\documentclass[preprint,aps,a4paper,showpacs,amsfonts,amsmath,amssymb]{revtex4-1}
\usepackage{makeidx}
\usepackage{graphicx}
\usepackage{epsfig}

\begin{document}

\title{First-passage and escape problems in the Feller process} 

\author{Jaume Masoliver}
\email{jaume.masoliver@ub.edu}
\affiliation{Departament de F\'{\i}sica Fonamental, Universitat de Barcelona,\\
Diagonal, 647, E-08028 Barcelona, Spain}
\author{Josep Perell\'o}
\email{josep.perello@ub.edu}
\affiliation{Departament de F\'{\i}sica Fonamental, Universitat de Barcelona,\\
Diagonal, 647, E-08028 Barcelona, Spain}

\date{\today}

\begin{abstract}

The Feller process is an one-dimensional diffusion process with linear drift and state-dependent diffusion coefficient vanishing at the origin. The process is positive definite and it is this property along with its linear character that have made Feller process a convenient candidate for the modeling of a number of phenomena ranging from single neuron firing to volatility of financial assets. While general properties of the process are well known since long, less known are properties related to level crossing such as the first-passage and the escape problems. In this work we thoroughly address these questions.

\end{abstract}
\maketitle

\section{Introduction}
\label{intro}

Diffusion processes are Markovian random processes with continuous sample paths. From a mathematical point of view they are characterized, in one dimension, by two functions: the drift, $f(y,t)$, and a positive defined diffusion coefficient $D(y,t)\geq 0$. The sample paths of any diffusion process can thus be pictured as the continuous trajectory resulting from the superposition of a deterministic evolution, governed by $f(y,t)$, and fluctuations around it, the latter determined by $D(y,t)$. Denoting the process by $Y(t)$, the diffusion picture becomes apparent by the fact that the time evolution of $Y(t)$ is ruled by the stochastic differential equation
$$
dY(t)=f(Y(t),t)dt+\sqrt{D(Y(t),t)}dW(t).
$$
where $W(t)$ is the Wiener process, that is, a Gaussian process with zero mean, unit variance and correlation function 
$\langle W(t_1)W(t_2)\rangle=\min(t_1,t_2)$. In what follows all stochastic differentials are interpreted in the sense of Ito. 

The Feller process is a special kind of diffusion process with linear drift and linear diffusion coefficient vanishing at the origin \cite{feller1}. The time evolution of the process is thus governed by
\begin{equation}
dY(t)=[-\alpha Y(t)+\beta]dt+k\sqrt{Y(t)}dW(t),
\label{sde1}
\end{equation}
where $\alpha>0$, $\beta$ and $k>0$ are constant parameters. 

Both Feller and Orstein-Uhlenbeck processes (diffusion processes also with linear drift but constant diffusion coefficient) have been widely used, with a marked prominence of the latter, in the modeling of countless physical phenomena. Both share a linear drift, $f(Y)=-\alpha Y+\beta$, which for $\alpha>0$ results in a restoring force that, in the absence of noise, makes both processes decay toward the value $\beta$. 

However, and contrary to the Ornstein-Uhlenbek process where diffusion is constant, the Feller process has a state-dependent diffusion coefficient, $D(Y)=k^2Y$, which for large values of $Y$ enhances the effects of noise while as $Y$ goes to zero the effect of noise vanishes. Hence, when the process reaches the origin, the drift drags it toward the value $\beta$. If $\beta>0$ the process, starting at some positive value, cannot reach the negative region which, in turn, renders the process always non-negative (otherwise the noise term in Eq. (\ref{sde1}) would become imaginary). Therefore, for the Feller process the origin is a singular boundary that the process cannot cross. 

A related question is whether or not the origin is accessible, in other words, whether the value $Y=0$ can or cannot be attained by the process. This is a crucial question in many practical situations and, as we will prove later, the answer depends on the particular value taken by a parameter which  balances the values of $\beta$ and $k^2$. The problem of classifying the different types of boundaries appearing in diffusion processes was thoroughly studied by Feller himself during the 1950's and we refer the reader to the literature for a more complete account on the subject \cite{feller2,feller3,gardiner}. 

Possessing a state-dependent diffusion and, most importantly, the fact that the process never attains negative values have made Feller process and ideal candidate for modeling a number of phenomena in natural and social sciences. Theoretical biology was one of the first places where the process was, during 1970's, seriously considered~\cite{ricciardibook}. Perhaps the most prominent place is within the context of neurobiology in order to model firing of single neurons~\cite{ricciardi,gerstner,lanska,lansky1995}. The Feller neuronal model is one of the so-called stochastic integrate and fire models which are simple models aiming to reproduce the membrane potential fluctuations. Experimental progresses has lead to the possibility of fitting real data to the Feller neuronal model among many others~\cite{lansky,ditlevsen,bibbona,jolivet,tchumatchenko}. 

In a different context Capocelli and Ricciardi~\cite{capocelli} considered the possibility of modeling biological population with the Feller process in order to include environmental randomness to the classic Malthusian growth rate~\cite{murray}. The approach~\cite{capocelli,azaele,azaelenature} represents in fact an alternative to the Lotka-Volterra models in ecosystems and the interest in this sort of problems is mostly focused on the extinction --that is, on the possibility of attaining the singular boundary $Y=0$-- and also on the unrestricted growth~\cite{azaele,azaelenature}. 

Financial markets is another field where the Feller process is widely used. It was introduced in 1985 to model term structure of interest rates receiving the name of Cox, Ingersoll, Ross (CIR) model and successfully evaluate bond prices~\cite{cox,hull}. The process is also being considered to provide a random character to the volatility of a given stock. Volatility obeying the Feller model jointly with a log-Brownian stochastic dynamics for the asset price evolution configure a two-dimensional diffusion market process called Heston model~\cite{heston,yakovenko} which is a rather useful model specially for option pricing~\cite{hull,heston}.

In all of the above mentioned situations susceptible to be modeled with a Feller process, the first-passage time events related, among others, to level-crossing and the triggering of a given signal are very significant phenomena for different reasons which depend on each context. This is for instance the case of the neuronal activity since a spike generation is due to the crossing of a threshold by the membrane potential signal. Additionally, population extinctions of any type or volatility bursts in financial markets are also important phenomena to model and study. First-passage time is, however, a difficult topic~\cite{redner,sokolov,klafter,shlesinger,weiss}. To our knowledge, for the Feller process this crucial facet has been scantily studied and only partially solved some years ago in the context of single neuron firing~\cite{ricciardi,lanska}. It is our main objective to address the first-passage time properties of process (\ref{sde1}). 

This paper is organized as follows. In Section~\ref{gen_prop}, we introduce the general properties of the unrestricted probability density of the Feller model. Section~\ref{fp} is devoted to the derivation of the first-passage time and escape probabilities with special attention to a couple of specific situations. Section~\ref{MFPT} is mostly focused on the derivation of the mean first-passage time. We finally summarize the results obtained in Section~\ref{summary}.

\section{General properties of the Feller model}
\label{gen_prop}

Before addressing the main issue of this paper, let us briefly review the main traits of the process and the role of the boundary at the origin. 

For the rest of the paper in turns out to be convenient to scale time and the process itself in the following way (recall we have assumed $\alpha>0$) 
\begin{equation}
t'=\alpha t, \qquad X=\frac{2\alpha}{k^2}Y,
\label{scale}
\end{equation}
so that the Langevin equation (\ref{sde1}) reads
\begin{equation}
dX(t')=-[X(t')-\theta]dt'+\sqrt{2X(t')}dW(t'),
\label{sde2}
\end{equation}
where $\theta$ is the only free parameter left. Its relation to $\beta$ and $k$ is 
\begin{equation}
\theta=\frac{2\beta}{k^2}>0.
\label{theta}
\end{equation}
This parameter is called ``saturation level'' or ``normal level'' and it is the value to which the Feller process $X(t')$ is attracted to. As we will shortly see, $\theta$ has a key role in the behavior of the process.

Let $p(x,t'|x_0)$ be the probability density function (pdf) for process (\ref{sde2}) to be in state $x$ at time $t'$:
$$
p(x,t'|x_0)dy={\rm Prob}\{x\leq X(t')<x+dx|X(0)=x_0\}.
$$
This density satisfies the (forward) Fokker-Planck equation (FPE) (as long as there is no confusion we will drop the prime in the time variable)
\begin{equation}
\frac{\partial p}{\partial t}=\frac{\partial}{\partial x}[(x-\theta)p]+\frac{\partial^2}{\partial x^2}(xp),
\label{fpe1}
\end{equation}
with initial condition
\begin{equation}
p(x,0|x_0)=\delta(x-x_0).
\label{initial1}
\end{equation}

Recall that $x=0$ is a singular boundary of the process and no ``particle'' can either leave or enter through this boundary (see Sect. \ref{intro}). A sufficient condition for this to happen is that the probability flux of the process through $x=0$ is zero \cite{gardiner}. We will thus search for solutions of the initial-value problem (\ref{fpe1})-(\ref{initial1}) that meet such a condition, that is, 
\begin{equation}
\lim_{x\rightarrow 0}\left\{(x-\theta)p(x,t|x_0)+\frac{\partial}{\partial x}[xp(x,t|x_0)]\right\}=0.
\label{flux}
\end{equation}

The expression for the pdf of the process $p(x,t|x_0)$ was first obtained by Feller himself many years ago using a tortuous procedure which involved the solution of a rather clumsy integral equation \cite{feller1}. In the Appendix \ref{appA} we present a simpler and more direct derivation based on the Laplace transform of the problem (\ref{fpe1})-(\ref{initial1}). The final expression reads
\begin{equation}
p(x,t|x_0)=\frac{1}{1-e^{-t}}\left(\frac{xe^{-t}}{x_0}\right)^{(\theta-1)/2}\exp\left\{-\frac{x+x_0e^{-t}}{1-e^{-t}}\right\}
I_{\theta-1}\left(\frac{2\sqrt{xx_0e^{-t}}}{1-e^{-t}}\right),
\label{final_pdf}
\end{equation}
where $I_{\theta-1}(z)$ is a modified Bessel function defined as \cite{mos}
\begin{equation}
I_{\theta-1}(z)=\sum_{n=0}^\infty\frac{(z/2)^{2n+\theta-1}}{n!\Gamma(n+\theta)}.
\label{bessel1}
\end{equation}

From Eq. (\ref{final_pdf}) we easily get the stationary pdf of the process defined as
$$
p_{\rm st}(x)=\lim_{t\rightarrow\infty}p(x,t|x_0).
$$
Indeed, taking into account that (cf. Eq. (\ref{bessel1})) 
\begin{equation}
I_{\theta-1}(z)=\frac{1}{\Gamma(\theta)}(z/2)^{\theta-1}\left[1+O(z^2)\right],
\label{bessel2}
\end{equation}
and from Eq. (\ref{final_pdf}) we obtain the Gamma distribution:
\begin{equation}
p_{\rm st}(x)=\frac{1}{\Gamma(\theta)}x^{\theta-1}e^{-x}.
\label{stationary}
\end{equation}

Another property that we can easily establish is the behavior of the probability distribution at the singular boundary located at $x=0$. In effect, using Eq. (\ref{bessel2}) we see from Eq. (\ref{final_pdf}) that
$$
p(x,t|x_0)=\frac{e^{-x_0e^{-t}/(1-e^{-t})}}{\Gamma(\theta)(1-e^{-t})^\theta}x^{\theta-1}\left[1+O(x)\right],
$$
from which it follows
\begin{equation}
p(0,t|x_0)= \begin{cases} \infty & \quad \theta<1 \\ 0 & \quad \theta>1,
\end{cases}
\label{boundary1}
\end{equation}
and
\begin{equation}
p(0,t|x_0)= \frac{e^{-x_0e^{-t}/(1-e^{-t})}}{1-e^{-t}} \qquad (\theta=1).
\label{boundary2}
\end{equation}

We thus see that when $\theta>1$ the probability for the Feller process to reach the value $x=0$ is zero but when $\theta\leq 1$ this probability is greater than zero. In other words, if $\theta\leq 1$ the origin is an accessible boundary, while if $\theta>1$ it is not \cite{feller1,feller2}. 

\section{First-passage and escape probabilities}
\label{fp}

After reviewing the main traits of the Feller process we will now focus on level-crossing problems --a collective name embracing questions such as hitting, first-passage, escape and extreme values, among others-- for that process. According to whether we are dealing with one-sided or two-sided barrier problems, we separate level crossing into two different issues. In one of them, the hitting or first-passage problem, we deal with the time that the process reaches some ``critical'' value, or ``threshold'', for the first time. The second issue, albeit closely related to the first one, concerns the time when the process first leaves a given interval. This is the so-called escape or exit problem. 

\subsection{First-passage probability}
\label{fp_a}

Let us first address the first-passage problem for the Feller process. The problem is solved when one knows the first-passage probability to threshold $x_c$. Let us denote by $W_c(t|x)$ the probability of first reaching $x_c\geq 0$ when the process starts at $t=0$ from the value $x>0$. 

As is well known \cite{gardiner,redner,weiss} the first-passage probability satisfies the backward Fokker-Planck equation
\begin{equation}
\frac{\partial W_c}{\partial t}=-(x-\theta)\frac{\partial W_c}{\partial x}+x\frac{\partial^2 W_c}{\partial x^2},
\label{bfpe1}
\end{equation}
with initial condition
\begin{equation}
W_c(0|x)=0,
\label{initial_1}
\end{equation}
and boundary condition
\begin{equation}
W_c(t|x_c)=1.
\label{bc1}
\end{equation}

The difficulty of solving the initial-boundary problem (\ref{bfpe1})-(\ref{bc1}) is decreased by taking the time Laplace transform,
$$
\hat W_c(s|x)=\int_0^\infty e^{-st}W_c(t|x)dt,
$$
which reduces the original problem to the solution of an ordinary differential equation (the Kummer equation \cite{mos}):
\begin{equation}
x\frac{d^2\hat W_c}{dx^2}-(x-\theta)\frac{d\hat W_c}{dx}-s\hat W_c=0,
\label{kummer}
\end{equation}
with boundary condition
\begin{equation}
\hat W_c(s|x_c)=\frac 1s.
\label{kummer_bc}
\end{equation}
Since $W_c$ is a probability it is obvious that any solution of the problem must be finite and non negative for all $x\geq 0$. 

The general solution of the Kummer equation (\ref{kummer}) is \cite{mos}
\begin{equation}
\hat W_c(s|x)=AF(s,\theta,x)+BU(s,\theta,x),
\label{general_sol}
\end{equation}
where $A$ and $B$ are arbitrary constants and $F(s,\theta,x)$ and $U(s,\theta,x)$ are the confluent hypergeometric functions of first and second kind \cite{mos} respectively defined by 
\begin{equation}
F(s,\theta,x)=\frac{\Gamma(\theta)}{\Gamma(s)}\sum_{n=0}^\infty\frac{\Gamma(s+n)}{\Gamma(\theta+n)}\frac{x^n}{n!}
\label{def_F}
\end{equation}
and
\begin{equation}
U(s,\theta,x)=\frac{\Gamma(1-\theta)}{\Gamma(s+1-\theta)} F(s,\theta,x)+\frac{\Gamma(\theta-1)}{\Gamma(s)}x^{\theta-1} F(s+1-\theta,2-\theta,x).
\label{def_U}
\end{equation}

In order to proceed further we need to specify whether the initial value $x$ is above or below the threshold $x_c$:

\subsubsection{Initial value below threshold ($x\leq x_c$)}

\begin{figure}[t]
\includegraphics[scale=0.65]{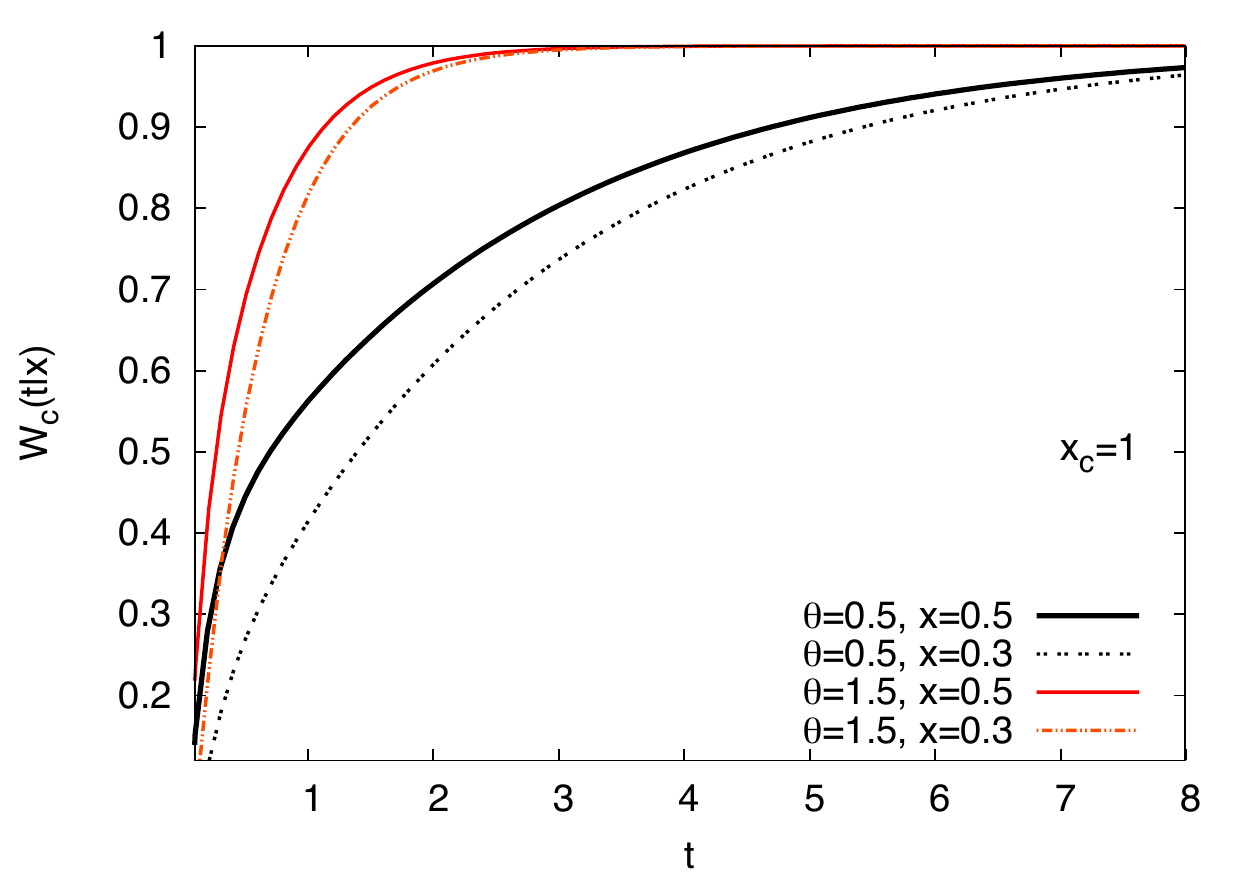}
\caption{First-passage probability $W_c(t|x)$ given by Eq.~(\ref{fpp_iib}) as a function of the scaled time and for two different parameters $\theta$ and initial values $x$. Inverse Laplace transform is obtained with the Stehfest algorithm.}\label{fig1}
\end{figure}

In this case $x$ can be arbitrarily small and taking into account that (see Eqs. (\ref{def_F})--(\ref{def_U})) 
\begin{equation}
\lim_{x\rightarrow 0} U(s,\theta,x)=\begin{cases}
\frac{\Gamma(1-\theta)}{\Gamma(s+1-\theta)} & \quad \theta<1 \\ \infty & \quad \theta> 1
\end{cases}
\label{u_0}
\end{equation}
we see that the solution to the problem staying finite for any initial position between the origin and $x_c$ and {\it for any positive value of the parameter $\theta$} is
$$
\hat W_c(s|x)=AF(s,\theta,x).
$$
The boundary condition (\ref{kummer_bc}) fixes the value of $A$ and
\begin{equation}
\hat W_c(s|x)=\frac{F(s,\theta,x)}{sF(s,\theta,x_c)}, \qquad (x\leq x_c).
\label{fpp_iib}
\end{equation} 
Figure~\ref{fig1} shows the numerical computation of this expression in the original Laplace space. We have used the well-known Stehfest algorithm~\cite{stehfest} and the results does not show any computational problem. As expected the closest to the threshold the fastest de first-passage time probability decays. And a greater $\theta$, corresponds to a smaller $W_c(t|x)$ as well.

\subsubsection{Initial value above threshold ($x\geq x_c$)}

In such a case $x$ can be arbitrarily large. Hence, taking into account that
$$
\lim_{x\rightarrow\infty} F(s,\theta,x)=\infty,
$$
while $U(s,\theta,x)$ stays finite for all positive values of $x$ \cite{mos}, we see from Eq. (\ref{general_sol}) that the general solution of the problem which remains finite for all $x>0$ is
$$
\hat W_c(s|x)=BU(s,\theta,x),
$$
and from the boundary condition (\ref{kummer_bc}) we conclude
\begin{equation}
\hat W_c(s|x)=\frac{U(s,\theta,x)}{sU(s,\theta,x_c)}, \qquad (x\geq x_c).
\label{fpp_i}
\end{equation}
Numerical inversion of this result is again easy to compute with standard algorithms. The small difference lies on the fact that the hypergeometric function of second kind~(\ref{def_U}) is slightly more complicated than the hypergeometric function of first kind~(\ref{def_F}).

\subsection{Reaching the origin}
\label{fp_b}

Another interesting quantity is the first-passage probability to threshold $x_c=0$; that is to say, the probability of first attaining the singular boundary of the process. This probability is relevant in the firing of neurons and it was addressed some years ago by by Capocelli and Ricciardi \cite{ricciardi} (see also the work of Laska et al \cite{lanska}) . Let us denote by $W_0$ the first-passage probability to the origin. Since our process is always positive, $x\geq 0$, we must use Eq. (\ref{fpp_i}) in order to evaluate $\hat W_0$. Setting $x_c=0$ in Eq. (\ref{fpp_i}) and using Eq. (\ref{u_0}) we obtain
\begin{equation}
\hat W_0(s|x)=
\begin{cases}
\frac{\Gamma(s+1-\theta)}{s\Gamma(1-\theta)}U(s,\theta,x) & \quad \theta<1 \\ 0 & \quad \theta> 1.
\end{cases}
\label{W_0}
\end{equation}

We will now proceed to invert Eq. (\ref{W_0}) thus obtaining the first-passage probability $W(t|x)$ in real time, something that seem to be unfeasible for any value of the threshold $x_c$, at least exactly (more on this below). 

Using the property \cite{mos}
$$
U(s,\theta,x)=x^{1-\theta}U(s+1-\theta,2-\theta,x),
$$
we write for $\theta<1$:
$$
\hat W_0(s|x)=\frac 1s \frac{\Gamma(s+1-\theta)}{\Gamma(1-\theta)} x^{1-\theta}U(s+1-\theta,2-\theta,x)^,
$$
which, after using the following integral representation of the Kummer function $U$ \cite{mos}
$$
U(a,c,x)=\frac{1}{\Gamma(a)}\int_0^\infty e^{-xz}z^{a-1}(1+z)^{c-a-1}dz,
$$
reads
$$
\hat W_0(s|x)=\frac{x^{1-\theta}}{s\Gamma(1-\theta)}\int_0^\infty e^{-xz}z^{-\theta}\left(\frac{z}{1+z}\right)^sdz.
$$

\begin{figure}[t]
\includegraphics[scale=0.65]{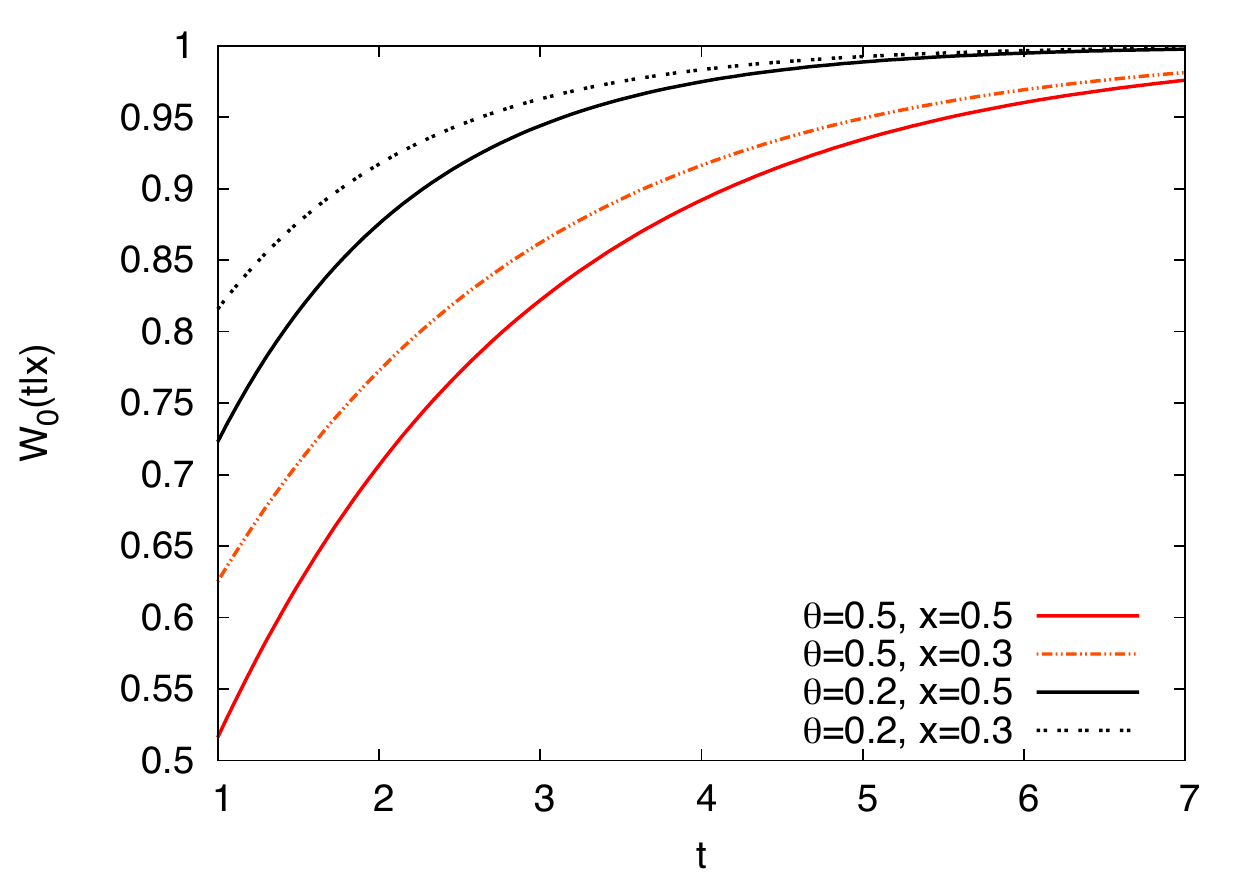}
\caption{First-passage probability $W_0(t|x)$ given by Eq.~(\ref{W_0_t}) as a function of time ($t>1)$ for two different values of $\theta$ ($\theta<1$) and the initial values $x$. Note that smaller values of $\theta$ and $x$ imply bigger first-passage probability.}
\label{fig2}
\end{figure}

Therefore,
$$
W_0(t|x)=\frac{x^{1-\theta}}{\Gamma(1-\theta)}\int_0^\infty e^{-xz}z^{-\theta}{\mathcal L}^{-1}\left\{\frac 1s\left(\frac{z}{1+z}\right)^s\right\}dz,
$$
where ${\mathcal L}^{-1}$ stands for Laplace inversion. Since \cite{roberts} 
\begin{equation}
{\mathcal L}^{-1}\left\{\frac{e^{-as}}{s}\right\}=\Theta(t-a)
\label{LI}
\end{equation}
where $\Theta(\cdot)$ is the Heaviside step function, we have
$$
{\mathcal L}^{-1}\left\{\frac 1s\left(\frac{z}{1+z}\right)^s\right\}=
\Theta\left[t+\ln\left(\frac{z}{1+z}\right)\right]=\Theta\left(z-\frac{e^{-t}}{1-e^{-t}}\right).
$$
Hence,
$$
W_0(t|x)=\frac{x^{1-\theta}}{\Gamma(1-\theta)}\int_{e^{-t}/(1-e^{-t})}^\infty e^{-xz}z^{-\theta}dz,
$$
or, equivalently,
$$
W_0(t|x)=\frac{1}{\Gamma(1-\theta)}\Gamma\left(1-\theta, \frac{xe^{-t}}{1-e^{-t}}\right),
$$
where $\Gamma(a,z)$ is the incomplete gamma function \cite{mos}
$$
\Gamma(a,z)=\int_z^\infty y^{a-1}e^{-y}dy.
$$
Finally,
\begin{equation}
W_0(t|x)=
\begin{cases}
\frac{1}{\Gamma(1-\theta)}\Gamma\left(1-\theta, \frac{xe^{-t}}{1-e^{-t}}\right) & \quad \theta<1 \\ 
0 & \quad \theta> 1.
\end{cases}
\label{W_0_t}
\end{equation}
We remark (as shown already in Eq. (\ref{W_0})) that when $\theta>1$ the first-passage probability to the origin is zero in agreement with the fact, pointed out in Sect. \ref{gen_prop}, that if $\theta>1$ $x=0$ is unattainable. The probability of first reaching the origin is represented in Fig.~\ref{fig2} where $W_0(t|x)$ is shown as a function of time and for two different values of parameter model $\theta$ and initial value $x$.

As $t\rightarrow\infty$ and for $\theta<1$ the first-passage probability $W_0$ becomes equal to $1$, as it is expected since if $\theta<1$ crossing the origin is a certain event as time grows. It is, however, interesting to see how $W_0$ approach the unity. To this end we use the following series expansion of the incomplete Gamma function \cite{mos}
$$
\Gamma(a,z)=\Gamma(a)-\sum_{n=0}^\infty \frac{(-1)^n}{n!}\frac{z^{a+n}}{a+n}.
$$
In the limit $t\rightarrow\infty$ we then get
$$
\Gamma\left(1-\theta, \frac{xe^{-t}}{1-e^{-t}}\right)=\Gamma(1-\theta)-\frac{1}{1-\theta}x^{1-\theta}e^{-(1-\theta)t}+O\left(e^{-(2-\theta)t}\right).
$$
Therefore,
\begin{equation}
W_0(t|x)=1-\frac{1}{\Gamma(2-\theta)}x^{1-\theta}e^{-(1-\theta)t}+O \left(e^{-(2-\theta)t}\right), \qquad (\theta<1). 
\label{W_o_assym}
\end{equation}
An interesting expression combining a power-law in $x$ and an exponential decay in time. 

\subsection{Large threshold}
\label{fp_c}

We will now study in some detail the interesting case of a large value of the threshold which is the opposite case considered above. It is clear that as $x_c\rightarrow\infty$ the threshold becomes unreachable and the first-passage probability approaches zero. Let us see how is the limiting process. These results can be a really interesting for instance when one wants to control financial asset volatility. A large volatility corresponds to wild fluctuations in the asset price.

In the case under consideration the initial position is always below threshold and the starting point of the analysis must be Eq. (\ref{fpp_iib}):
$$
\hat W_c(s|x)=\frac{F(s,\theta,x)}{sF(s,\theta,x_c)}.
$$
Since now $x_c\rightarrow\infty$ we use the following asymptotic expansion of the Kummer function $F$ \cite{mos}
$$
F(s,\theta,x_c)=\frac{\Gamma(\theta)}{\Gamma(s)}e^{x_c}x_c^{s-\theta}\left[1+O\left(\frac{1}{x_c}\right)\right],
$$
and
$$
\hat W_c(s|x)=\frac{\Gamma(s)}{s\Gamma(\theta)}e^{x_c}x_c^{s-\theta}F(s,\theta,x)\left[1+O\left(\frac{1}{x_c}\right)\right].
$$

Our next step is the use of the following integral representation of $F$ \cite{mos}
$$
F(s,\theta,x)=\frac{1}{\Gamma(s)}\int_0^\infty e^{-z}z^{s-1}F(\theta,xz) dz,
$$
where $F(\theta,xz)$ is the following hypergeometric series \cite{mos}
\begin{equation}
F(\theta,xz)=\sum_{n=0}^\infty\frac{1}{(\theta)_n}\frac{(xz)^n}{n!}.
\label{hyper}
\end{equation}
Hence, for large values of the threshold, we approximately have
$$
\hat W_c(s|x)\simeq \frac{1}{s\Gamma(\theta)}e^{-x_c}x_c^{\theta-s}\int_0^\infty e^{-z}z^{s-1}F(\theta,xz) dz, \qquad(x_c\rightarrow\infty).
$$

We now proceed as in Sec. \ref{fp_b}. The Laplace inversion of the last equation reads
$$
W_c(t|x)\simeq \frac{1}{\Gamma(\theta)}e^{-x_c}x_c^{\theta}\int_0^\infty\frac{e^{-z}}{z}F(\theta,xz)\mathcal{L}^{-1}\left\{\frac 1s\left(\frac{z}{x_c}\right)^s\right\}dz,
\qquad(x_c\rightarrow\infty). 
$$
Since 
$$
\left(\frac{z}{x_c}\right)^s=\exp\left[s\ln\left(\frac{z}{x_c}\right)\right],
$$
then, recalling Eq. (\ref{LI}), we have
$$
\mathcal{L}^{-1}\left\{\frac 1s\left(\frac{z}{x_c}\right)^s\right\}=\Theta\left[t+\ln\left(\frac{z}{x_c}\right)\right]=\Theta\left(z-x_ce^{-t}\right).
$$
Therefore,
\begin{equation}
W_c(t|x)\simeq \frac{1}{\Gamma(\theta)}x_c^{\theta}e^{-x_c}\int_{x_ce^{-t}}^\infty\frac{e^{-z}}{z}F(\theta,xz) dz, \qquad(x_c\rightarrow\infty).
\label{lx_1}
\end{equation}

Using Eq. (\ref{hyper}) we can give an alternative expression to Eq. (\ref{lx_1}) which is somewhat more convenient for numerical work. It reads
\begin{equation}
W_c(t|x)\simeq \frac{1}{\Gamma(\theta)}x_c^{\theta}e^{-x_c}\sum_{n=0}^\infty \frac{1}{(\theta)_n}\frac{x^n}{n!}\Gamma(n,x_ce^{-t}), \qquad(x_c\rightarrow\infty),
\label{lx_2}
\end{equation}
where $\Gamma(n,x_ce^{-t})$ is the incomplete Gamma function. This expression is particularly suited for small values of the initial position. Thus, for instance, when $x=0$ we write 
\begin{equation}
W_c(t|0)\simeq \frac{1}{\Gamma(\theta)}x_c^{\theta}e^{-x_c}E_1\left(x_ce^{-t}\right), \qquad(x_c\rightarrow\infty),
\label{lx_3}
\end{equation}
where 
$$
E_1(x)=\int_{x}^\infty\frac{e^{-z}}{z}dz,
$$
is the exponential integral. 

\subsection{The escape probability}
\label{fp_d}

We close this section by briefly addressing the escape problem which, as mentioned before, is closely related with the first-passage problem studied above. The problem at hand consists in knowing whether or not the process $X(t)$, starting at some point inside an interval $(a,b)$, has left this interval for the first time. The answer lies in the knowledge of the survival probability $S_{ab}(t|x)$ defined as the probability that, starting at $x\in (a,b)$, the process at time $t$ has not left the interval at that time or during any previous instant of time: 
$$
S_{ab}(t|x)={\rm Prob}\bigl\{a<X(t')<b;\ 0\leq t'\leq t\ |\ a<x<b\bigr\},
$$
where $x=X(0)$ is the starting point. The escape probability, {\it i.e.}, the probability that at time $t$ the process has exited the interval $(a,b)$ for the first time, is then given by
$$
W_{ab}(t|x)=1-S_{ab}(t|x).
$$

As is well known \cite{gardiner,redner,weiss} the survival probability obeys the backward Fokker-Planck equation 
$$
\frac{\partial S_{ab}}{\partial t}=-(x-\theta)\frac{\partial S_{ab}}{\partial x}+x\frac{\partial^2 S_{ab}}{\partial x^2},
$$
with initial and boundary conditions given by
$$
S_{ab}(0|x)=1, \qquad S_{ab}(t|a)=S_{ab}(t|b)=0.
$$
Hence, the escape probability is the solution of the initial and boundary value problem (compare with Eqs. (\ref{bfpe1})--(\ref{bc1}))
\begin{equation}
\frac{\partial W_{ab}}{\partial t}=-(x-\theta)\frac{\partial W_{ab}}{\partial x}+x\frac{\partial^2 W_{ab}}{\partial x^2},
\label{bfpe2}
\end{equation}
\begin{equation}
W_{ab}(0|x)=0, \qquad W_{ab}(t|a)=W_{ab}(t|b)=1.
\label{initial-bc}
\end{equation}

Following the same reasoning as before (see Eqs. (\ref{kummer}) and (\ref{kummer_bc})) we see that the time Laplace transform of the escape probability $\hat W(s|x)$ satisfies the boundary value problem
\begin{equation}
x\frac{d^2\hat W_{ab}}{dx^2}-(x-\theta)\frac{d\hat W_{ab}}{dx}-s\hat W_{ab}=0,
\label{kummer_2}
\end{equation}
\begin{equation}
\hat W_{ab}(s|a)=\hat W_{ab}(s|b)=\frac 1s.
\label{kummer_bc_2}
\end{equation}
Again, The general solution of the Kummer equation (\ref{kummer_2}) is \cite{mos}
$$
\hat W_{ab}(s|x)=AF(s,\theta,x)+BU(s,\theta,x),
$$
where $A$ and $B$ are arbitrary constants and $F$ and $U$ are defined in Eqs. (\ref{def_F}) and (\ref{def_U}). 

Boundary conditions (\ref{kummer_bc_2}) determine the value of $A$ and $B$ and, after routine algebra, the final result for the escape probability reads
\begin{equation}
\hat W_{ab}(s|x)=\frac{\bigl[U(s,\theta,b)-U(s,\theta,a)\bigr]F(s,\theta,x)-\bigl[F(s,\theta,b)-F(s,\theta,a)\bigr]U(s,\theta,x)}
{s\bigl[F(s,\theta,a)U(s,\theta,b)-F(s,\theta,b)U(s,\theta,a)\bigr]},
\label{exit_prob}
\end{equation}
$(a\leq x\leq b)$. 

\section{Long-time asymptotic behavior and mean first-passage times}
\label{MFPT}

In the previous section we have solved the hitting and escape problems for the Feller process by means of the evaluation of the first-passage and exit probabilities. We have obtained exact analytical expressions for the time-Laplace transform of these probabilities. Unfortunately exact inversion seems to be beyond reach except for the cases in Sec. \ref{fp_b}--when the threshold is located at the origin-- and in the following Sec. \ref{MFPT_a} --with approximate expressions suitable for long times. In this section we will also obtain two important magnitudes associated with the problem: the mean first-passage time, $T_c(x)$, and the mean escape time, $T_{ab}(x)$.

\subsection{Long-time behavior of the first-passage probability}
\label{MFPT_a}

Let $\tau_c(x)$ be the first-passage time for the process, starting at $x$, to reach some threshold $x_c$ for the first time. It is a random variable depending on each realization of the process. Formally
$$
\tau_c(x)={\rm inf}\bigl\{t|X(t)>x_c; X(0)=x<x_c\bigr\}
$$
when the initial value is below threshold, and
$$
\tau_c(x)={\rm inf}\bigl\{t|X(t)<x_c; X(0)=x>x_c\bigr\}
$$
when the initial value is above threshold.

We next relate the first-passage time with the hitting probability $W_c(t|x)$ defined in the previous section. Note that if $\tau_c(x)$ is the first-passage time, the hitting probability can be defined as
$$
W_c(t|x)={\rm Prob}\{\tau_c(x)\leq t\}
$$
which shows that $W_c(t|x)$ is the distribution function of the first-passage time. The corresponding probability density is thus defined 
$$
f_c(t|x)dt={\rm Prob}\{t\leq\tau_c(x)<t+dt\},
$$
and it is related to the distribution $W_c$ by
\begin{equation}
f_c(t|x)=\frac{\partial W_c(t|x)}{\partial t}.
\label{fW}
\end{equation}

The moments of this distribution are
$$
T_n(x|x_c)=\int_0^\infty t^nf_c(t|x)dt,
$$
$(n=1,2,3,\dots)$ and the mean first-passage time (MFPT) is the first moment: 
$$
T_c(x)\equiv T_1(x|x_c).
$$

Note that in terms of the Laplace transform
$$
\hat f_c(s|x)=\int_0^\infty e^{-st}f_c(t|x)dt,
$$
the first-passage moments are
$$
T_n(x|x_c)=(-1)^n\left.\frac{\partial^n\hat f_c(s|x)}{\partial s^n}\right|_{s=0},
$$
which implies that, as long as $T_n(x|x_c)$ ($n=1,2,3,\dots$) exist, the Laplace transform of the first-passage time density has the following expansion in powers of $s$
\begin{equation}
\hat f_c(s|x)=\sum_{n=0}^\infty\frac{(-1)^n}{n!}T_n(x|x_c)s^n.
\label{expan_1}
\end{equation}
On the other hand the Laplace transform of Eq. (\ref{fW}), along with the initial condition $W_c(0|x)=0$, yields
\begin{equation}
\hat W_c(s|x)=\frac 1s \hat f(s|x).
\label{w_f}
\end{equation}
By combining Eqs. (\ref{expan_1}) and (\ref{w_f}) we then have 
\begin{equation}
\hat W_c(s|x)=\sum_{n=0}^\infty\frac{(-1)^n}{n!}s^{n-1}T_n(x|x_c),
\label{expan_2}
\end{equation}
expansion that furnishes the basis for the asymptotic analysis of the first-passage probability $W(t|x)$. Indeed, the so-called Tauberian theorems prove that the long-time behavior of a function $g(t)$ is determined by the small $s$ behavior of its Laplace transform $\hat g(s)$ \cite{tauberian}. For the case of the first-passage probability we see from 
Eq. (\ref{expan_2}) that the small $s$ behavior of $\hat W_c$ is 
\begin{equation}
\hat W_c(s|x)=\frac 1s-T_c(x)+O(s)=\frac 1s[1-sT_c(x)+O(s^2)].
\label{small_s}
\end{equation}
where $T_c(x)\equiv T_1(x|x_c)$ is the mean first-passage time. Note that expansion (\ref{small_s}) may also be written, within the same level of approximation, as 
\begin{equation}
\hat W_c(s|x)=\frac{1}{s[1+sT_c(x)+O(s^2)]},
\label{expan_3}
\end{equation}
which by the Tauberian theorems \cite{tauberian} implies that the long-time behavior of the first-passage probability $W_c(t|x)$ is given by the Laplace inversion of 
Eq. (\ref{expan_3}). That is,
\begin{equation}
W_c(t|x)\simeq 1-e^{-t/T_c(x)}, \qquad (t\rightarrow\infty). 
\label{asym_1}
\end{equation}

We have thus obtained the long-time behavior of the first-passage probability to threshold $x_c$ and see that the MFPT determines the long-time behavior of the first-passage probability. We will next evaluate this average time for the Feller process.

\subsection{The mean first-passage time}
\label{MFPT_b}

In terms of the Laplace transform of first-passage probability $\hat W_c(s|x)$ obtaining the MFPT is straightforward. In effect from Eq. (\ref{small_s}) we see that
\begin{equation}
T_c(x)=\lim_{s\rightarrow 0}\left[\frac 1s - \hat W_c(s|x)\right].
\label{T_c_def}
\end{equation}
Using the findings of Sec. \ref{fp} we know that the first-passage probability has different expressions as to whether the initial value of the process $x$ is above or below the threshold $x_c$. Let us now treat these two cases including the special case $x_c=0$. 

\subsubsection{Initial value above threshold ($x\geq x_c$)}

In this case (cf Eq. (\ref{fpp_i})) 
$$
\hat W_c(s|x)=\frac{U(s,\theta,x)}{sU(s,\theta,x_c)},
$$
and 
\begin{equation}
T_c(x)=\lim_{s\rightarrow 0}\left[\frac 1s \frac{U(s,\theta,x_c)-U(s,\theta,x)}{U(s,\theta,x_c)}\right].
\label{T_a}
\end{equation}
The expansion in powers of $s$ of the Kummer function $U(s,\theta,x)$ is presented in Appendix \ref{appB} where it is shown that
\begin{equation}
U(s,\theta,x)=1+sU_1(x)+O(s^2),
\label{u_expan}
\end{equation}
where
\begin{equation}
U_1(x)\equiv-\psi(1-\theta)-\int^x U(1,1+\theta,z)dz,
\label{u_1}
\end{equation}
and $\psi(z)=\Gamma'(z)/\Gamma(z)$ is the psi function. We note that the function, defined as the indefinite integral: 
$$
\int^x U(1,1+\theta,z) dz
$$
cannot be reduced to another Kummer function \cite{mos} nor, to the best of our knowledge, to any other tabulated function. 

Plugging Eqs. (\ref{u_expan})-(\ref{u_1}) into Eq. (\ref{T_a}) we finally obtain
\begin{equation}
T_c(x)=\int_{x_c}^x U(1,1+\theta,z)dz, \qquad (x\geq x_c).
\label{T_b}
\end{equation}

\subsubsection{MFPT to the origin}

In Sects. \ref{gen_prop} and \ref{fp} (see Eq. (\ref{W_0})) we have seen that when $\theta>1$ the origin is unattainable. However, if $\theta<1$ this singular boundary can be reached by the process. In this later case it is natural to ask which is the MFPT to the origin. The question has not only an academic interest but is relevant in mathematical biology where $x=0$ corresponds to the potential at which a neuron is fired \cite{ricciardi}. Also in econophysics it is useful to know whether volatility or the interest rates can drop to zero and which is the average time expected to do so. 

Since $x>0$, the expression for the MFPT to the origin, denoted by $T_0(x)$, will be given by Eq. (\ref{T_b}) with $x_c=0$. Unfortunately setting $x_c=0$ in Eq. (\ref{T_b}) is not possible because the integral is singular at the lower level. 

We proceed as follows: start with the definition of the Kummer function $U$ given in Eq. (\ref{def_U}), use the integration rule \cite{mos}
$$
\int^x F(a,c,z)dz=\frac{x^c}{c}F(a,c+1,x),
$$
and take into account the standard property of the Gamma function $\Gamma(z+1)=z\Gamma(z)$. We write
\begin{equation}
\int^x U(1,1+\theta,z)dz=-\frac 1\theta \int^x F(1,1+\theta,z)dz-\Gamma(\theta-1)x^{1-\theta}F(1-\theta,2-\theta,x).
\label{int_u}
\end{equation}
Substituting into Eq. (\ref{T_b}) and taking the limit $x_c\rightarrow 0^+$, we have
\begin{eqnarray*}
T_0(x)&=&\lim_{x_c\rightarrow 0^+}\Biggl\{-\frac 1\theta\int_{x_c}^x F(1,1+\theta,z) dz\\
&-&\Gamma(\theta-1)\Bigl[x^{1-\theta}F(1-\theta,2-\theta,x)-x_c^{1-\theta}F(1-\theta,2-\theta,x_c)\Bigr]\Biggr\}.
\end{eqnarray*}
Using the value of the Kummer function $F$ at the origin, $F(1,1+\theta,0)=1$ (cf Eq. (\ref{def_F})), we have
\begin{eqnarray}
T_0(x)&=&-\frac 1\theta\int_{0}^x F(1,1+\theta,z) dz \nonumber \\ 
&-&\Gamma(\theta-1)\left[x^{1-\theta}F(1-\theta,2-\theta,x)-\lim_{x_c\rightarrow 0^+}\left(x^{1-\theta}\right)\right].
\label{T0_int}
\end{eqnarray}
Hence, if $\theta<1$ we get
$$
T_0(x)=-\frac 1\theta\int_{0}^x F(1,1+\theta,z) dz-\Gamma(\theta-1)x^{1-\theta}F(1-\theta,2-\theta,x),
$$
and taking into account Eq. (\ref{int_u}) we see that in this case the expression for the MFPT to the origin is given by Eq. (\ref{T_b}) with $x_c=0$:
\begin{equation}
T_0(x)=\int_0^x U(1,1+\theta,z)dz \qquad (\theta<1).
\label{T_0-}
\end{equation}
However, we see from Eq. (\ref{T0_int}) that when $\theta>1$ $x_c^{1-\theta}\rightarrow\infty$ as $x_c\rightarrow 0^+$ and the process takes an infinite average time to reach the origin
\begin{equation}
T_0(x)=\infty \qquad (\theta>1),
\label{T_0+}
\end{equation}
which confirms that when $\theta>1$ the singular boundary $x=0$ is unattainable.

\begin{figure}[t]
\includegraphics[scale=0.7]{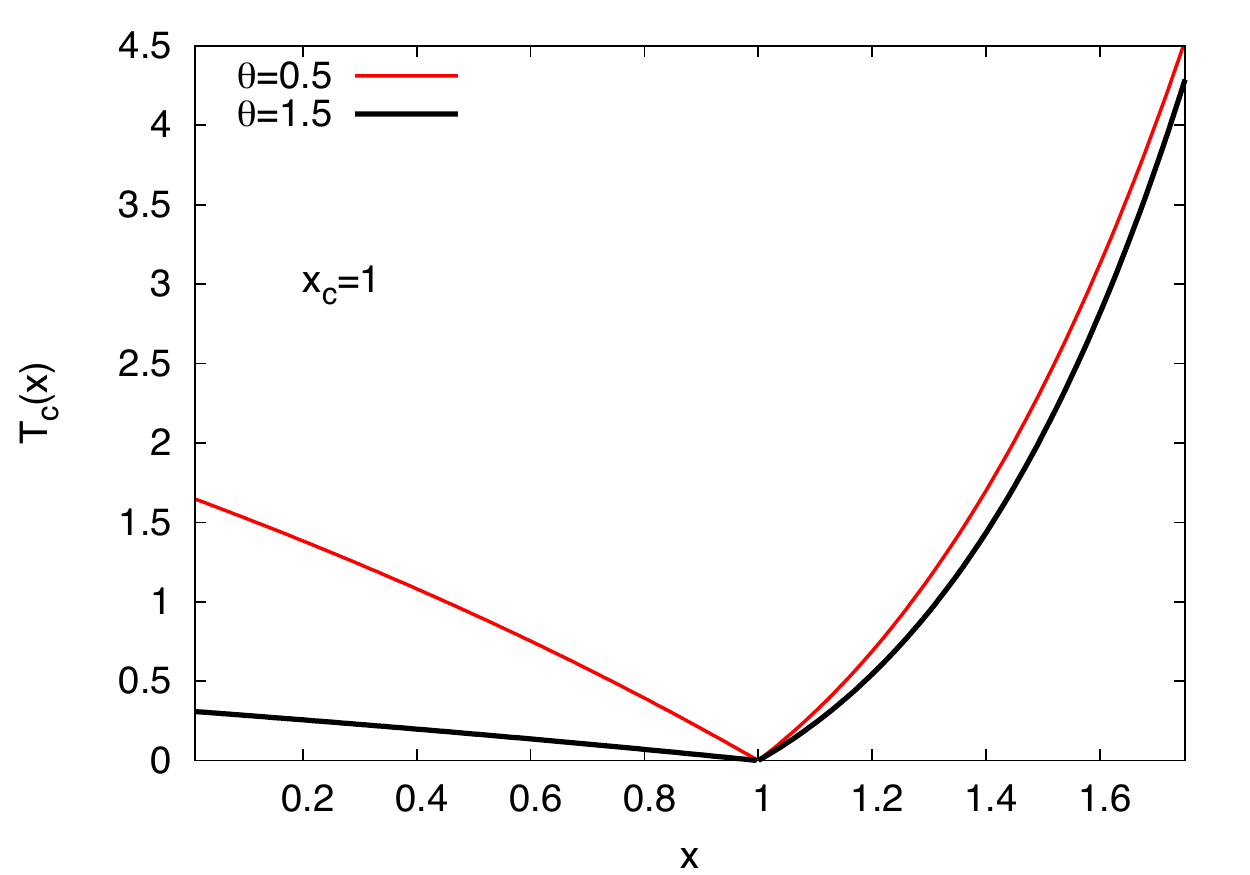}
\caption{The mean first-passage time as function of the initial value $x$ for two different values of $\theta$ when the critical value $x_c$ equals to one. The figure jointly shows all MFPT~(\ref{T_b_i}) when $x<x_c$ and the MFPT~(\ref{T_b}) when $x>x_c$. Putting together the two mathematical expressions allow us to clearly observe the distinct and asymmetric behavior of the mean first-passage depending on whether $x<x_c$ or $x>x_c$.}\label{fig3}
\end{figure}

\subsubsection{Initial value below threshold ($x\leq x_c$)}

Now (see Eq. (\ref{fpp_iib}))
$$
\hat W_c(s|x)=\frac{F(s,\theta,x)}{sF(s,\theta,x_c)}. 
$$
Hence, 
\begin{equation}
T_c(x)=\lim_{s\rightarrow 0}\left[\frac 1s \frac{F(s,\theta,x_c)-F(s,\theta,x)}{F(s,\theta,x_c)}\right].
\label{T_b_lim}
\end{equation}
In the Appendix \ref{appB} we show that the expansion of powers of $s$ of the Kummer function $F(s,\theta,x)$ is
\begin{equation}
F(s,\theta,x)=1+sF_1(x)+O(s^2),
\label{F_expan}
\end{equation}
where 
\begin{equation}
F_1(x)\equiv\frac 1\theta \int_0^x F(1,1+\theta,z)dz.
\label{F_1}
\end{equation}
Substituting Eqs. (\ref{F_expan})-(\ref{F_1}) into Eq. (\ref{T_b_lim}) yields
\begin{equation}
T_c(x)=\frac 1\theta \int_{x}^{x_c} F(1,1+\theta,z)dz, \qquad (x\leq x_c).
\label{T_b_i}
\end{equation}
A result we obtained few years ago \cite{mp07} in another context and using a different approach. This result that applies for $x\leq x_c$ jointly with previous Eq.~(\ref{T_b}) that applies for $x\geq x_c$ are presented in Fig.~\ref{fig3}. We show there the marked asymmetric behavior of the MFPT depending on whether the initial value $x$ is larger or smaller than the critical value $x_c$

\subsection{The mean escape time}
\label{MET}

We close this section by obtaining the time taken by the process $X(t)$ starting at $X(0)=x$ to first leave a given interval $(a,b)$, where $a<x<b$. This is called the escape (or exit) time out of an interval, $\tau_{ab}(x)$, and is formally defines as
$$
\tau_{ab}(x)={\rm inf}\bigl\{t | a\geq X(t) \geq b; a<x<b \bigr\}.
$$
The exit time is a random variable characterized by a distribution function,
$$
{\rm Prob}\bigl\{\tau_{ab}(x)<t|X(0)=x\bigr\}={\rm Prob}\bigl\{a\geq X(t) \geq b|X(0)=x\bigr\},
$$
which is precisely the escape probability $W_{ab}(t|x)$ discussed in Sec. \ref{fp}. The moments of the exit time are thus defined by
$$
T_{ab}^{(n)}(x)=\int_0^\infty t^ndW_{ab}(t|x),
$$
($n=1,2,3,\dots$) and the mean escape time (MET) is the first moment:
$$
T_{ab}^{(1)}(x)\equiv T_{ab}(x).
$$

Proceeding as in Sec. \ref{MFPT_a} we easily see that the Laplace transform of the escape probability can be written as (cf Eq. (\ref{expan_2})) 
$$
\hat W_{ab}(s|x)=\sum_{n=0}^\infty\frac{(-1)^n}{n!}s^{n-1}T_{ab}^{(n)}(x),
$$
from which it follows that (cf Eq. (\ref{T_c_def}))
\begin{equation}
T_{ab}(x)=\lim_{s\rightarrow 0}\left[\frac 1s - \hat W_{ab}(s|x)\right].
\label{T_ab}
\end{equation}
Moreover, similarly to the first-passage problem discussed above, we can easily prove that the long-time behavior of the escape probability is also solely determined by the MET (see Eq. (\ref{asym_1})): 
\begin{equation}
W_{ab}(t|x)\simeq 1-e^{-t/T_{ab}(x)}, \qquad (t\rightarrow\infty).
\label{asym_2}
\end{equation}

Plugging the expression for $\hat W_{ab}(s|x)$ given by Eq. (\ref{exit_prob}) into Eq. (\ref{T_ab}) and taking into account the small $s$ development of the Kummer functions $U$ and $F$, as expressed respectively by Eqs. (\ref{u_expan})-(\ref{u_1}) and Eqs. (\ref{F_expan})-(\ref{F_1}) we obtain after lengthy but otherwise straightforward algebra the following expression of the MET:
\begin{equation}
T_{ab}(x)=\frac{N_{ab}(x)}{D_{ab}(x)},
\label{T_exit}
\end{equation}
where
\begin{eqnarray}
N_{ab}(x)&=&\int_a^xU(1,1+\theta,z)dz \int_0^{b}F(1,1+\theta,z)dz \nonumber \\
&+&\int_x^bU(1,1+\theta,z)dz \int_0^{a}F(1,1+\theta,z)dz \nonumber \\ 
&-&\int_a^{b}U(1,1+\theta,z)dz \int_0^xF(1,1+\theta,z)dz,
\label{N}
\end{eqnarray}
and 
\begin{equation}
D_{ab}(x)=\int_a^{b}\bigl[F(1,1+\theta,z)+\theta U(1,1+\theta,z)\bigr]dz.
\label{D}
\end{equation}

\section{Summary and Conclusions}
\label{summary}

We have fully addressed the first-passage and escape problems for the Feller process. Let us now summarize the main results obtained. The process is an one-dimensional diffusion defined by a linear drift and a linear diffusion coefficient vanishing at the origin. Feller process has the property of being positive, a salient characteristic which has earned the process some popularity in modeling several phenomena, from neural activity to financial markets. 

Perhaps the best way to define the process is by means of a stochastic differential equation. In the dimensionless units defined in Eq. (\ref{scale}) of Sec. \ref{gen_prop} this reads 
$$
dX(t)=-[X(t)-\theta]dt+\sqrt{2X(t)}dW(t),
$$
(we have dropped the prime in the time variable) where $W(t)$ is the Wiener process and $\theta>0$ is the saturation or normal level to which $X(t)$ is attracted to as $t$ increases. The origin is a singular boundary because the noise term vanishes there. In Sec. \ref{gen_prop} we have reviewed the general properties of the processes which were mostly obtained by Feller many years ago. One of these properties refer to the attainability of the origin in which the normal level plays a crucial role. Thus if $\theta\leq 1$ the origin is an accessible boundary while if $\theta>1$ it is not. 

The bulk of the paper is developed in Secs. \ref{fp} and \ref{MFPT} where the first-passage and escape properties of the Feller process are thoroughly analyzed. The first-passage problem refers to the crossing by the process of certain preassigned critical level or threshold $x_c$ while the escape problem refers to the departure of some interval $(a,b)$. 

\bigskip

The first-passage properties are fully characterized by the hitting, or first-passage probability, defined as the probability of first reaching the threshold $x_c$ by the process at time $t$ or before. We denote by $W_c(t|x)$ this probability, where $x$ is the initial position of the process. This probability depends on whether the process is initially below ($x\leq x_c$) or above ($x\geq x_c$) threshold. We have obtained exact expressions for the Laplace transform of the hitting probability,
$$
\hat W_c(s|x)=\int_0^\infty e^{-st}W_c(t|x)dt,
$$
which are summarized as
$$
\hat W_c(s|x)= \begin{cases} \frac{F(s,\theta,x)}{sF(s,\theta,x_c)}, & \quad x\leq x_c, \\ 
\frac{U(s,\theta,x)}{sU(s,\theta,x_c)}, & \quad x\geq x_c,
\end{cases}
$$
where $F$ and $U$ are Kummer functions. 

In general these expressions for the Laplace transform of the hitting probability cannot be inverted exactly in an analytical fashion and one has to resort to the numerical inversion. There are some instances, however, in which we have been able to obtain analytical expressions in real time. This is the case of hitting the origin which has a significant interest in the firing of neurons and also in the Heston volatility model of the financial analysis. We denote by $W_0(t|x)$ the first-passage probability to threshold $x_c=0$, we have shown that
$$
W_0(t|x)=\begin{cases}
\frac{1}{\Gamma(1-\theta)}\Gamma\left(1-\theta,\frac{xe^{-t}}{1-e^{-t}}\right), & \quad \theta<1, \\
0, & \quad \theta>1,
\end{cases}
$$ 
where $\Gamma(a,z)$ is the incomplete gamma function. If $\theta<1$ $W_0(t|x)\rightarrow 1$ as $t\rightarrow\infty$. This the expected behavior since when $\theta<1$ crossing the origin is a sure event as time grows. The way $W_0$ approaches unity is explicitly given by the following combination of a power-law in the initial position and an exponential time decay
$$
W_0(t|x)=1-\frac{1}{\Gamma(2-\theta)}x^{1-\theta}e^{-(1-\theta)t}+O(e^{-(2-\theta)t}), \qquad (\theta<1).
$$

Another instance in which we have been able to obtain an (approximate) expression for the first-passage probability in real time is when threshold is large. In such a case
$$
W_c(t|x)\simeq \frac{1}{\Gamma(\theta)}x_c^{\theta}e^{-x_c}\sum_{n=0}^\infty \frac{1}{(\theta)_n}\frac{x^n}{n!}\Gamma(n,x_ce^{-t}), \qquad(x_c\rightarrow\infty),
$$
where $\Gamma(n,x_ce^{-t})$ is the incomplete Gamma function. 

\bigskip

The escape problem is completely characterized by the escape probability, $W_{ab}(t|x)$, defined as the probability of first leaving a given interval $(a,b)$. It is complementary to the survival probability $S_{ab}$:
$$
W_{ab}(t|x)=1-S_{ab}(t|x),
$$
where $S_{ab}(t|x)$ is the probability that the process has not exited $(a,b)$ at time $t$ or during any previous instant of time. Formally,
$$
S_{ab}(t|x)={\rm Prob}\bigl\{X(t')\in(a,b),\ 0\leq t'\leq t\ |X(0)=x\in(a,b)\ \bigr\}.
$$

For the Feller process we have been able to obtain the exact expression for the Laplace transform of the escape probability which turns out to be more involved than the first-passage probability. It reads
$$
\hat W_{ab}(s|x)=\frac{\bigl[U(s,\theta,b)-U(s,\theta,a)\bigr]F(s,\theta,x)-\bigl[F(s,\theta,b)-F(s,\theta,a)\bigr]U(s,\theta,x)}
{s\bigl[F(s,\theta,a)U(s,\theta,b)-F(s,\theta,b)U(s,\theta,a)\bigr]},
$$
where $a\leq x\leq b$, and $F$ and $U$ are Kummer functions.

\bigskip

We have next addressed the problem of the mean-first passage time (MFPT) and the mean exit time (MET). We have shown that in terms of the first-passage time moments $T_n(x|x_c)$ ($n=1,2,3,\dots$) --of which the MFPT corresponds to $n=1$, $T_1(x|x_c)\equiv T_c(x$)-- the Laplace transform of the hitting probability reads
$$
\hat W_c(s|x)=\frac 1s+\sum_{n=1}^\infty\frac{(-1)^n}{n!}s^{n-1}T_n(x|x_c).
$$
The MFPT to the threshold $x_c$ is then given by
$$
T_c(x)=\lim_{s\rightarrow 0}\left[\frac 1s-\hat W_c(s|x)\right].
$$
From these expressions we have been able to obtain, in terms of the MFPT, the following long-time asymptotic expression of the hitting probability
$$
W_c(t|x)\simeq 1-e^{-t/T_c(x)}, \qquad (t\rightarrow\infty).
$$

For the Feller process this analysis has led to different results according to whether initially the system is placed below or above the threshold:
$$
T_c(x)=\begin{cases} 
(1/\theta)\int_{x_c}^x F(1,1+\theta,z)dz, & \quad x\leq x_c, \\
\int_{x_c}^x U(1,1+\theta,z)dz, & \quad x\geq x_c.
\end{cases}
$$
The MFPT to reach the origin, $T_0(x)$, has also been analyzed with the result
$$
T_0(x)=
\begin{cases}
\int_0^x U(1,1+\theta,z)dz, & \quad \theta<1,\\ 
\infty, & \quad \theta>1,
\end{cases}
$$
which constitute an additional proof of the fact that when $\theta>1$ the singular boundary $x=0$ is unattainable. 

The analysis of the MFPT can be exactly carried out for the MET. The resulting expressions relating the MET with the escape probability are formally the same as those relating the MFPT with the hitting probability as can be seen in Sec. \ref{MET}. Thus, for instance,
$$
T_{ab}(x)=\lim_{s\rightarrow 0}\left[\frac 1s-\hat W_{ab}(s|x)\right],
$$
and 
$$
W_{ab}(t|x)\simeq 1-e^{-t/T_{ab}(x)}, \qquad (t\rightarrow\infty),
$$
where $W_{ab}(t|x)$ and $T_{ab}(x)$ are the escape probability and the MET respectively. In the case of the Feller process the explicit expression for the MET is given in Eqs. (\ref{T_exit})-(\ref{D}). 

Let us finally mention that the extension of the above results to the study of the extreme values attained by the process, such as the maximum and minimum values, as well as their application to financial time series --in particular the volatility-- is under present research and we expect getting a number of results very soon. 

\acknowledgments

Partial financial support from the Ministerio de Ciencia e Innovaci\'{o}n under Contract No. FIS 2009-09689 is acknowledged.

\appendix
 
\section{The probability density function}
\label{appA}

The solution to problem (\ref{fpe1})-(\ref{initial1}) is more conveniently addressed by its Laplace transform with respect to $x$:
\begin{equation}
\hat{p}(\sigma,t|x_0)=\int_0^\infty e^{-\sigma x}p(x,t|x_0) dx.
\label{LT}
\end{equation}
Taking into account condition (\ref{flux}), the transformed problem (\ref{fpe1})-(\ref{initial1}) reads
\begin{equation}
\frac{\partial\hat p}{\partial t}+\sigma(1+\sigma)\frac{\partial\hat p}{\partial\sigma}=-\theta\sigma\hat p,
\label{fpe2}
\end{equation}
\begin{equation}
\hat p(\sigma,0|x_0)=e^{-\sigma x_0}.
\label{initial2}
\end{equation} 

Equation (\ref{fpe2}) is a linear partial differential equation of first order whose solution can be obtained by the method of characteristics \cite{courant}. In effect, let the function $h(\sigma)$ be defined by the characteristic of Eq. (\ref{fpe2}), $h'(\sigma)=-[\sigma(1+\sigma)]^{-1}$, that is
\begin{equation}
h(\sigma)=\ln\left(\frac{1+\sigma}{\sigma}\right).
\label{h}
\end{equation}
Then the solution of Eq. (\ref{fpe2}), as can be rightly seen by direct substitution, is \cite{courant}
\begin{equation}
\hat p(\sigma,t|x_0)=(1+\sigma)^{-\theta}\psi(t+h(\sigma)),
\label{intermediate_1}
\end{equation}
where $\psi(z)$ is an arbitrary function to be determined by the initial condition (\ref{initial2}), {\it i.e.,}
$$
\psi(h(\sigma))=(1+\sigma)^\theta e^{-\sigma x_0},
$$
which implies, after inverting Eq. (\ref{h}) to write $\sigma$ in terms of $h$, that
$$
\psi(z)=(1-e^{-z})^{-\theta}\exp\left\{-\frac{x_0e^{-z}}{1-e^{-z}}\right\}.
$$
Substituting this into Eq. (\ref{intermediate_1}) we finally obtain
\begin{equation}
\hat p(\sigma,t|x_0)=\frac{1}{[1+\sigma(1-e^{-t})]^\theta}\exp\left\{-\frac{\sigma x_0e^{-t}}{1+\sigma(1-e^{-t})}\right\}.
\label{A1}
\end{equation}

Let us now proceed to the Laplace inversion of Eq. (\ref{A1}). Calling
\begin{equation}
a=1-e^{-t}, \qquad b=x_0e^{-t},
\label{A2}
\end{equation}
simple algebraic manipulations followed by a power expansion yield
$$
\exp\left\{-\frac{b\sigma}{1+a\sigma}\right\}=
e^{-b/a}\exp\left\{-\frac{b}{a(1+a\sigma)}\right\}=e^{-b/a}\sum_{n=0}^\infty\frac{(b/a)^n}{n!(1+a\sigma)^n}.
$$
Plugging into Eq. (\ref{A1}) we get
\begin{equation}
\hat p(\sigma,t|x_0)=e^{-b/a}\sum_{n=0}^\infty\frac{(b/a)^n}{n!(1+a\sigma)^{n+\theta}}.
\label{A3}
\end{equation}

Let us denote by $\mathcal{L}^{-1}\{\hat f(\sigma)\}=f(x)$ the operation of Laplace inverting $\hat f(\sigma)$ and recall the standard property 
$$
\mathcal{L}^{-1}\{\hat f(a\sigma+1)\}=\frac 1a e^{-x/a} f(x/a)
$$
and also 
$$
\mathcal{L}^{-1}\left\{\frac{1}{\sigma^{n+\theta}}\right\}=\frac{x^{n+\theta-1}}{\Gamma(n+\theta)}.
$$
Then the Laplace inversion of Eq. (\ref{A3}) yields
$$
p(x,t|x_0)=\frac 1a e^{-(x+b)/a}\sum_{n=0}^\infty\frac{(b/a)^n(x/a)^{n+\theta-1}}{n!\Gamma(n+\theta)}
$$
which after simple manipulations reads
$$
p(x,t|x_0)=\frac 1a\left(\sqrt{\frac xb}\right)^{\theta-1}e^{-(x+b)/a}\sum_{n=0}^\infty\frac{(\sqrt{bx}/a)^{2n+\theta-1}}{n!\Gamma(n+\theta)}.
$$

We recognize the series appearing in this equation as the expression of a modified Bessel Function. Indeed 
$$
I_\nu(z)=\sum_{n=0}^\infty\frac{(z/2)^{2n+\nu}}{n!\Gamma(n+\nu+1)},
$$
where $I_\nu(z)$ is the modified Bessel function of order $\nu$ \cite{mos}. Therefore, 
$$
p(x,t|x_0)=\frac 1a\left(\frac xb\right)^{(\theta-1)/2}e^{-(x+b)/a}I_{\theta-1}\left(\frac{2\sqrt{bx}}{a}\right),
$$
which after reverting to the original notation (cf. Eq. (\ref{A2})) reads
$$
p(x,t|x_0)=\frac{1}{1-e^{-t}}\left(\frac{xe^{-t}}{x_0}\right)^{(\theta-1)/2}\exp\left\{-\frac{x+x_0e^{-t}}{1-e^{-t}}\right\}
I_{\theta-1}\left(\frac{2\sqrt{xx_0e^{-t}}}{1-e^{-t}}\right),
$$
which is Eq. (\ref{final_pdf}).

\section{Expansions for $F(s,\theta,x)$ and $U(s,\theta,x)$}
\label{appB}

In terms of the Pochhammer's symbol $(a)_n=\Gamma(a+n)/\Gamma(a)$ Kummer function $F$ is defined as the series \cite{mos}
$$
F(s,\theta,x)=\sum_{n=0}^\infty\frac{(s)_n}{(\theta)_n}\frac{x^n}{n!}.
$$
Since $(s)_0=1$ and 
$$
(s)_n=s(s+1)(s+2)\cdots(s+n-1)=s(n-1)!+O(s^2)
$$
we have
\begin{equation}
F(s,\theta,x)=1+s\sum_{n=1}^\infty\frac{1}{(\theta)_n}\frac{x^n}{n}+O(s^2).
\label{b1}
\end{equation}
In the sum of the right hand side we make the replacement $n\rightarrow n+1$ and take into account that $(\theta)_{n+1}=\theta(\theta+1)_n$, we thus write
$$
\sum_{n=1}^\infty\frac{1}{(\theta)_n}\frac{x^n}{n}=\frac 1\theta\sum_{n=0}^\infty\frac{1}{(\theta+1)_{n}}\frac{x^{n+1}}{n+1}=
\frac 1\theta\sum_{n=0}^\infty\frac{1}{(\theta+1)_{n}}\int_0^x z^ndz.
$$
We can easily see that $(1)_n=n!$, hence
$$
\sum_{n=1}^\infty\frac{1}{(\theta)_n}\frac{x^n}{n}=\frac 1\theta\int_0^x\left[\sum_{n=0}^\infty\frac{(1)_n}{(\theta+1)_{n}}\frac{z^n}{n!}\right]dz,
$$
which, after recalling the definition of the confluent hypergeometric function $F$, Eq. (\ref{def_F}), yields
$$
\sum_{n=1}^\infty\frac{1}{(\theta)_n}\frac{x^n}{n}=\frac 1\theta\int_0^x F(1,1+\theta,z)dz.
$$
Substituting into Eq. (\ref{b1}) we get
\begin{equation}
F(s,\theta,x)=1+sF_1(x)+O(s^2),
\label{b2}
\end{equation}
where
\begin{equation}
F_1(x)\equiv \frac 1\theta \int_0^x F(1,1+\theta,z)dz.
\label{b3}
\end{equation}

The small $s$ expansion of the Kummer function of second kind $U(s,\theta,x)$ is a bit more involved. We start from the definition of $U$ in terms of $F$ (cf Eq. (\ref{def_U})) 
\begin{equation}
U(s,\theta,x)=\frac{\Gamma(1-\theta)}{\Gamma(1-\theta+s)}F(s,\theta,x)+\frac{\Gamma(\theta-1)}{\Gamma(s)}x^{\theta-1}F(1-\theta+s,2-\theta,x),
\label{b4}
\end{equation}
then expand 
\begin{eqnarray}
\Gamma(1-\theta+s)&=&\Gamma(1-\theta)+s\Gamma'(1-\theta)+O(s^2) \nonumber\\
&=&\Gamma(1-\theta)\Bigl[1+s\psi(1-\theta)+O(s^2)\Bigr],
\label{b5}
\end{eqnarray}
where $\psi(z)=\Gamma'(z)/\Gamma(z)$ is the psi function \cite{mos}. Also \cite{abramovich} 
\begin{equation}
\Gamma(s)=\frac 1s\left[1-\gamma s+O(s^2)\right],
\label{b6}
\end{equation}
where $\gamma=0.5772\cdots$ is Euler's constant. Plugging Eqs. (\ref{b2}), (\ref{b5}) and (\ref{b6}) into Eq. (\ref{b4}) we get
\begin{equation}
U(s,\theta,x)=1+sU_1(x)+O(s^2),
\label{b7}
\end{equation}
where
\begin{equation}
U_1(x)\equiv F_1(x)-\psi(1-\theta)+\Gamma(\theta-1)F(1-\theta,2-\theta,x).
\label{b8}
\end{equation}

Let us finally show that a more convenient form for $U_1(x)$ is given by
\begin{equation}
U_1(x)=-\psi(1-\theta)-\int^x U(1,1+\theta,z)dz..
\label{b9}
\end{equation}
In effect, recalling the definition of $F_1(x)$ given in Eq. (\ref{b3}) and using the integration rule $\int x^{c-1}F(a,c,x)dx=(x^c/c)F(a,c+1,x)$, we have 
$$
U_1(x)=-\psi(1-\theta)-\int\left[\frac{\Gamma(-\theta)}{\Gamma(1-\theta)}F(1,1+\theta,x)+\Gamma(\theta)x^{\theta}F(1-\theta,1-\theta,x)\right]dx,
$$
where we have used the well known property $\Gamma(1+z)=z\Gamma(z)$ to write $1/\theta=-\Gamma(-\theta)/\Gamma(1-\theta)$ and $(1-\theta)\Gamma(\theta-1)=-\Gamma(\theta)$. Note that the integrand is precisely the Kummer function of second kind 
$$
U(1,1+\theta,x)=\frac{\Gamma(-\theta)}{\Gamma(1-\theta)}F(1,1+\theta,x)+\Gamma(\theta)x^{\theta}F(1-\theta,1-\theta,x)
$$
(see Eq. (\ref{b4}) with $s=1$ and $\theta$ replaced by $1+\theta$). We have thus proved Eq. (\ref{b9}).

\end{document}